# Students' Voices on Generative AI: Perceptions, Benefits, and Challenges in Higher Education

Cecilia Ka Yuk CHAN[1*1], Wenjie HU[2]

*Corresponding Author

**Affiliation:** The University of Hong Kong
**Address:** Centre for the Enhancement of Teaching and Learning (CETL), Room CPD-1.81, Centennial Campus, The University of Hong Kong, Pokfulam, Hong Kong
Email: Cecilia.Chan@cetl.hku.hk [1]
Email: carol_hwj@connect.hku.hk [2]
Website: https://tlerg.cetl.hku.hk/

## Abstract

This study explores university students' perceptions of generative AI (GenAI) technologies, such as ChatGPT, in higher education, focusing on familiarity, their willingness to engage, potential benefits and challenges, and effective integration. A survey of 399 undergraduate and postgraduate students from various disciplines in Hong Kong revealed a generally positive attitude towards GenAI in teaching and learning. Students recognized the potential for personalized learning support, writing and brainstorming assistance, and research and analysis capabilities. However, concerns about accuracy, privacy, ethical issues, and the impact on personal development, career prospects, and societal values were also expressed. According to John Biggs' 3P model, student perceptions significantly influence learning approaches and outcomes. By understanding students' perceptions, educators and policymakers can tailor GenAI technologies to address needs and concerns while promoting effective learning outcomes. Insights from this study can inform policy development around the integration of GenAI technologies into higher education. By understanding students' perceptions and addressing their concerns, policymakers can create well-informed guidelines and strategies for the responsible and effective implementation of GenAI tools, ultimately enhancing teaching and learning experiences in higher education.

## Generative Artifical Intelligence

Generative AI (GenAI) encompasses a group of machine learning algorithms designed to generate new data samples that mimic existing datasets. One of the foundational techniques in GenAI is the Variational Autoencoder (VAE), which is a type of neural network that learns to encode and decode data in a way that maintains its essential features (Kingma & Welling, 2013). Another popular GenAI method is Generative Adversarial Networks (GANs), which consist of two neural networks working in competition to generate realistic data samples (Goodfellow et al., 2014). GenAI models use advanced algorithms to learn patterns and generate new content such as text, images, sounds, videos, and code. Some examples of GenAI tools include

ChatGPT, Bard, Stable Diffusion, and Dall-E. Its ability to handle complex prompts and produce human-like output has led to research and interest into the integration of GenAI in various fields such as healthcare, medicine, education, media, and tourism.

ChatGPT, for example, has caused a surge of interest in the use of GenAI in higher education since its release in November 2022 (Hu, 2023). It is a conversational AI system developed by OpenAI, an autoregressive large language model (more than 175 billion parameters) has been pre-trained on a large corpus of text data. It can generate human-like responses to a wide range of text-based inputs. The model has been trained on a diverse range of texts, including books, articles, and websites, allowing it to understand user input, generate responses, and maintain coherent conversations on a wide range of topics. There has been much discussion on its potential in transforming disciplinary practices such as medical writing (Biswas, 2023; Kitamura, 2023), surgical practice (Bhattacharya et al., 2023), and health care communications (Eggmann et al., 2023) as well as enhancing higher education teaching and learning (e.g., Adiguzel et al., 2023; Baidoo-Anu & Ansah, 2023).

**Benefits and challenges of using generative AI in higher education**
One of the key uses of GenAI in higher education is for enhancing students' learning experience through its ability to respond to user prompts to generate highly original output. Text-to-text AI generators can provide writing assistance to students, especially non-native students, by enabling them to brainstorm ideas and get feedback on their writing through applications such as ChatGPT (Atlas, 2023), while text-to-image AI generators such as DALL-E and Stable Diffusion can serve as valuable tools for teaching technical and artistic concepts in arts and design (Dehouche & Dehouche, 2023). GenAI tools are also believed to be useful research aids for generating ideas, synthesizing information, and summarising a vast amount of text data to help researchers analyse data and compose their writing (Berg, 2023), contributing to efficiency in publication (Kitamura, 2023; van Dis et al., 2023). Another opportunity in which GenAI can bring benefits is learning assessment (Crompton & Burke, 2023). Tools such as the Intelligent Essay Assessor are used to grade students' written work and provide feedback on their performance (Landauer, 2003). Mizumoto and Eguchi (2023) examined the reliability and accuracy of ChatGPT as an automated essay scoring tool, and the results show that ChatGPT shortened the time needed for grading, ensured consistency in scoring, and was able to provide immediate scores and feedback on students' writing skills. Such research demonstrates that GenAI has potential to transform the teaching and learning process as well as improve student outcomes in higher education.

On the other hand, there have been challenges about the limitations of GenAI and issues related to ethics, plagiarism, and academic integrity. Kumar's (2023) analysis of AI-generated responses to academic writing prompts shows that the text output, although mostly original and relevant to the topics, contained inappropriate references and lacked personal perspectives that AI is generally incapable of producing. For second language learners, constructing appropriate prompts poses a challenge in itself as it requires a certain level of linguistic skills; and overreliance on GenAI tools may compromise students' genuine efforts to develop writing competence (Warschauer et al., 2023). In addition, the content produced by GenAI may be biased, inaccurate, or harmful if the dataset on which a model was trained contains such elements (Harrer, 2023). AI-generated images, for example, may contain nudity or obscenity and can be created for malicious purposes such as deepfakes (Maerten & Soydaner, 2023). GenAI tools are not able to assess validity of content and determine whether the output they generate contains falsehoods or misinformation, thus their use requires human oversight (Lubowitz, 2023). Furthermore, since AI-generated output cannot be detected by most plagiarism checkers, it is difficult to determine whether a given piece of writing is the author's original work (Peres et al., 2023). As Zhai (2022) cautions, the use of text-to-text generators such as ChatGPT may compromise the validity of assessment practices, particularly those involving written assignments. Hence, the widespread use of GenAI can pose a serious threat to academic integrity in higher education. The benefits of GenAI underline the potential of the technology as a valuable learning tool for students, while its limitations and challenges show a need for research into how GenAI can be effectively integrated in the teaching and learning process. Thus, the research questions for this study are

1. How familiar are university students with GenAI technologies like ChatGPT?
2. What are the potential benefits and challenges associated with using GenAI in teaching and learning, as perceived by university students?
3. How can GenAI be effectively integrated into higher education to enhance teaching and learning outcomes?

**Student perceptions of the use of GenAI in higher education**

User acceptance is key to successful uptake of technological innovations (Davis, 1989). John Biggs emphasized the importance of student perception in his 3P(Presage-Process-Product) model of teaching and learning (Biggs, 2011). According to Biggs, students' perceptions of their learning environment, their abilities, and the teaching strategies used have a significant impact on their approach to learning (Biggs, 1999), which in turn influences their learning outcomes. Students who perceive the learning environment (such as, curriculum content, teaching methods, assessment methods, learning resources, learning context, student support services) positively and feel confident about their abilities are more likely to adopt a deep approach to learning, which involves seeking understanding and making connections between concepts. On the other hand, students who have a negative perception of their learning environment or doubt their abilities may adopt a surface approach to learning, where they focus on memorizing facts and meeting minimum requirements (Biggs, 2011). In a learning environment, the way students perceive a technological innovation such as GenAI, their views, concerns, and experiences of the technology can have impact on their willingness to utilise the tool and consequently the extent to which the tool is integrated in the learning process. A large proportion of research into tertiary students' perceptions in this area focuses on AI in general and chatbots which are not necessarily powered by GenAI, while students' views and experiences of GenAI tools specifically remain relatively underexplored. Research into student perceptions of AI/GenAI typically investigates students' attitudes, their experiences of AI, and factors influencing their perceptions such as gender, disciplines, age, and year of study.

***Attitudes towards AI and experiences of AI.*** Research into the use of AI in language classrooms shows that students found AI tools such as chatbots and Plot Generator useful for enhancing language acquisition by providing assistance with grammar, guiding them in generating ideas, and helping them communicate in the target language (Bailey et al., 2021; Sumakul et al., 2022). AI KAKU, a GenAI tool based on the GPT-2 language model, was implemented in English language lessons with Japanese students and was perceived to be easy to use and able to assist students to express themselves in English (Gayed et al., 2022); while the use of AI-based chatbots for learning support improved students' learning achievement, self-efficacy, learning attitude, and learning motivation (Essel et al., 2022; Lee et al., 2022). A study of the use of chatbots in business education also reported favourable user feedback with students citing positive learning experience due to chatbots' responsiveness, interactivity, and confidential learning support (Chen et al., 2023). Most students agreed that AI have a profound impact on their disciplines and future careers (e.g., Bisdas et al., 2021; Gong et al., 2019; Sit et al., 2020) and expressed an intention to utilise AI in their learning and future practice (e.g., Bisdas et al., 2021; Lee et al., 2022), and thus viewed integration of AI as an essential part of university curricula (e.g., Abdelwahab et al., 2023; Bisdas et al., 2021; Gong et al., 2019; Yüzbaşioğlu, 2021).

Students who had a good understanding of AI were also found to express a low level of anxiety about AI in Dahmash et al.'s (2020) study. However, Jeffrey's (2020) study found conflicting beliefs among college students. Students who had a high level of understanding and information about AI and believed that AI could benefit them personally also expressed concerns about the impact of AI on human jobs. In Dahmash et al.'s (2020) and Gong et al.'s (2019) research, the choice of radiology as a future career was associated with the impact of AI—The number of medical students indicating radiology as their specialty choice increased when the potential impact of AI was not a consideration. Among the concerns and drawbacks regarding the use of AI, as perceived by students, are limited human interaction/element (e.g., Bisdas et al., 2021; Essel et al., 2022), potential data leakage (e.g., Bisdas et al., 2021), absence of emotional connection (Chen et al., 2023), breach of ethics (e.g., Gillissen et al., 2022; Jha et al., 2022), and reduced job opportunities or increased demand in job practices (Gong et al., 2019; Ghotbi et al., 2022; Park et al., 2020).

***Frequency of use/Time spent on AI tools.*** Research examining the relationship between frequency of AI use and student perceptions of AI is inconclusive. For example, Yildiz Durak's (2023) study of 86 students in a university in Turkey reported no correlation between chatbot usage frequency and visual design self-efficacy, course satisfaction, chatbot usage satisfaction, and learner autonomy. The finding shows that frequency of use alone is not a meaningful factor, while satisfaction with use can impact users' self-efficacy. In contrast, Bailey et al. (2021) found that the amount of time spent on chatbot use in a second language writing class was positively associated with students' confidence in using the target language and perception of task value.

***Use of Methodology.*** Most of the research into student perceptions of AI/GenAI employs a quantitative survey design (e.g., Bisdas et al., 2021; Dahmash et al., 2020; Gherhes & Obrad, 2018; Yüzbaşioğlu, 2021). Some studies incorporated open-ended survey questions (e.g., Hew et al., 2023; Jeffrey, 2020) and semi-structured interviews (e.g., Gillissen et al., 2022; Mokmin & Ibrahim, 2021; Park et al., 2020) to gather students' free responses and to probe their views on the research topic in addition to their responses to survey questions. For example, Park et al.'s (2020) study consisted of two stages: Semi-structured interviews were conducted face-to-face or by telephone in Stage 1, followed by an Internet-based survey in Stage 2. Studies that examined the impact of AI and student perceptions typically adopted an experimental design using a pretest-intervention-posttest approach and the administration of a questionnaire to examine student perceptions (e.g., Essel et al., 2022; Lee et al., 2022). Qualitative research is relatively rare as only the views of a small number of students can be explored with such an approach. For example, Sumakul et al.'s (2022) and Terblanche et al.'s (2022) studies based on semi-structured interviews involved eight students and 20 students respectively. In contrast, survey is more effective for reaching a large population of respondents from different geographical locations as shown in previous studies such as Bisdas et al. (2021), Dahmash et al., (2020), and Gong et al. (2019).

Although there has been a considerable amount of research into AI in general as shown in the review of current studies in this section, there is currently lack of investigation into how students perceive GenAI. In view of the unprecedented interest in GenAI at present, there is a need to examine university students' attitude towards GenAI and their experience of using GenAI in order to gain insights into how it can be integrated in higher education to enhance teaching and learning.

**Methodology**

In this study, we used a survey design to collect data from university students in Hong Kong, exploring their use and perceptions of GenAI in teaching and learning. The survey was administered via an online questionnaire, consisting of both closed-ended and open-ended questions in order a large population of responses. Topics covered in the survey encompassed their knowledge of GenAI technologies like ChatGPT, the incorporation of AI technologies in higher education, potential challenges related to AI technologies, and the influence of AI on teaching and learning.

Data were gathered through an online survey, targeting students from all post-secondary educational institutions to ensure that the results represented the needs and values of all participants. A convenience sampling method was employed to select respondents based on their availability and willingness to partake in the study. Participants were recruited through an online platform and given an informed consent form before completing the survey.

A total of 399 undergraduate and postgraduate students, from various disciplines in Hong Kong, completed the survey. Descriptive analysis was utilized to analyze the survey data, and a thematic analysis approach was applied to examine the responses from the open-ended questions in the survey.

**Results**

**Demographic information**

Participants in this study were from ten faculties (Faculty of Architecture, Arts, Business, Dentistry, Education, Engineering, Law, Medicine, Science and Social Sciences), comprising 204 males (51.1%) and 195 females (48.9%). There were (44.4%, n = 177) undergraduate students and (55.6%, n = 222) postgraduate students. Nearly half of them (55.4%, n = 221) were enrolled in STEM fields, mainly from the Faculty of Engineering (33.1%) and the Faculty of Science (14.5%), while non-STEM students were primarily majored in Arts (14.8%, n=59), Business(13.3%, n=53) and Education(7.5%, n=30). Additionally, 66.7% participants have reported using GenAI technologies at least once. Specifically, 21.8% reported rarely using it, 29.1% using it sometimes, 9.8% often using it, and 6.0% reported always using it. Table 1 shows the demographics information.

| Characteristic | n | % |
|---|---|---|
| Sex | | |
|    Male | 204 | 51.1% |
|    Female | 195 | 48.9% |
| Academic level | | |
|    Undergraduate | 177 | 44.4% |
|    Postgraduate | 222 | 55.6% |
| Major | | |
|    STEM | 221 | 55.4% |
|    Non-STEM | 173 | 43.4% |
| Have you ever used generative AI technologies like ChatGPT? | | |
|    Never | 133 | 33.3% |
|    Rarely | 87 | 21.8% |
|    Sometimes | 116 | 29.1% |
|    Often | 39 | 9.8% |
|    Always | 24 | 6.0% |

Table 1 Demographic Information

**Knowledge of Generative AI Technologies**

As illustrated in Table 2, participants had a generally good understanding of GenAI technologies, with mean scores ranging from 3.89 to 4.15. Specifically, students had the highest mean score for the statement "I understand generative AI technologies like ChatGPT have limitations in their ability to handle complex tasks" (Mean=4.15, SD=0.82) and the lowest mean score for the emotional intelligence and empathy considerations(Mean=3.89, SD=0.97), indicating that while they generally understand GenAI technologies has limitations, they may not be fully aware of the potential risks arise from the lack of emotional intelligence and empathy.

Moreover, the data showed a moderate positive correlation between their knowledge of GenAI technologies and frequency of use(r=0.1, p<0.05). Specifically, regarding their agreement on if GenAI technologies like ChatGPT may generate factually inaccurate output, students who never or rarely use GenAI technologies (Mean=3.99, SD=0.847) were significantly different (t=2.695, p<0.01) from students who have used them at least sometimes (Mean=4.22 SD=0.829).

| Statement | Mean | SD |
|---|---|---|
| I understand generative AI technologies like ChatGPT have limitations in their ability to handle complex tasks. | 4.15 | 0.82 |
| I understand generative AI technologies like ChatGPT can generate output that is factually inaccurate. | 4.10 | 0.85 |
| I understand generative AI technologies like ChatGPT can generate output that is out of context or inappropriate. | 4.03 | 0.83 |
| I understand generative AI technologies like ChatGPT can exhibit biases and unfairness in their output. | 3.93 | 0.92 |
| I understand generative AI technologies like ChatGPT may rely too heavily on statistics, which can limit their usefulness in certain contexts. | 3.93 | 0.93 |
| I understand generative AI technologies like ChatGPT have limited emotional intelligence and empathy, which can lead to output that is insensitive or inappropriate. | 3.89 | 0.97 |

Table 2 Knowledge of Generative AI Technologies

**Willingness to use Generative AI Technologies**

Overall, the findings suggest that students have a positive attitude toward GenAI technologies. They would like to integrate GenAI technologies like ChatGPT in their learning practices (Mean=3.85, SD=1.02), as well as future careers (Mean=4.05; SD=0.96). Specifically, students highly value its perceived usefulness in providing unique insights (Mean=3.74; SD=1.08) and personalized feedback (Mean=3.61; SD=1.06). Additionally, they find these technologies are user-friendly, as they are available 24/7(Mean=4.12; SD=0.83) and offer anonymous support services (Mean=3.77; SD=0.99).

Moreover, the correlation analysis results show that students' perceived willingness to use GenAI technologies is positively correlated with both knowledge of GenAI (r=0.189; p<0.001) and frequency of use (r=0.326; p<0.001), indicating that students who are more knowledgeable about these technologies and use them more frequently are more likely to use them in the future.

| Statement | Mean | SD |
|---|---|---|
| I envision integrating generative AI technologies like ChatGPT into my teaching and learning practices in the future. | 3.85 | 1.02 |
| Students must learn how to use generative AI technologies well for their careers. | 4.05 | 0.96 |
| I believe generative AI technologies such as ChatGPT can improve my digital competence. | 3.70 | 0.96 |
| I believe generative AI technologies such as ChatGPT can help me save time. | 4.20 | 0.82 |
| I believe AI technologies such as ChatGPT can provide me with unique insights and perspectives that I may not have thought of myself. | 3.74 | 1.08 |
| I think AI technologies such as ChatGPT can provide me with personalized and immediate feedback and suggestions for my assignments. | 3.61 | 1.06 |
| I think AI technologies such as ChatGPT is a great tool as it is available 24/7. | 4.12 | 0.83 |

| | Mean | SD |
|---|---|---|
| I think AI technologies such as ChatGPT is a great tool for student support services due to anonymity. | 3.77 | 0.99 |

<p align="center">Table 3 Willingness to use Generative AI Technologies</p>

## Concerns about Generative AI Technologies

Unlike willingness, descriptive statistics show that students expressed a slight favor of concerns about GenAI. They expressed the least positive opinions about if people will become over-reliant on GenAI technologies (Mean=2.89; SD=1.13), and the highest rating was for how these technologies could affect the value of university education (Mean=3.18; SD=1.16).

Interestingly, there were significant differences between students who never or rarely used these technologies and other participants ($t=3.873$, $p<0.01$). However, no significant correlation was found between students' concerns and knowledge about GenAI technologies ($r=0.096$; $p>0.05$).

| Statement | Mean | SD |
|---|---|---|
| Using generative AI technologies such as ChatGPT to complete assignments undermines the value of university education. | 3.15 | 1.17 |
| Generative AI technologies such as ChatGPT will limit my opportunities to interact with others and socialize while completing coursework. | 3.06 | 1.20 |
| Generative AI technologies such as ChatGPT will hinder my development of generic or transferable skills such as teamwork, problem-solving, and leadership skills. | 3.10 | 1.23 |
| I can become over-reliant on generative AI technologies | 2.85 | 1.13 |

<p align="center">Table 4 Concerns about Generative AI Technologies</p>

## The Benefits and Challenges for Students' Willingness and Concerns

### *What are the reasons behind students' willingness to utilise generative AI technologies?*

Consistent with the findings from the quantitative data, most participants perceived GenAI as a valuable tool with numerous benefits and were willing to work with it, primarily on learning, writing and research purposes:

### (1) Personalized and immediate learning support

When students struggle with assignments, GenAI can act as a virtual tutor, providing personalized learning support and answering their questions immediately. A student from the faculty of engineering considered AI as "*a top student*" in their class, because "*When I have doubt and couldn't find other people to help me out, ChatGPT seems like a good option*." Besides immediate answers, customized recommendations and feedback were also valued by students. As one remarked, "*It would be useful if ChatGPT could help me find the most effective solution when I am checking my finished homework. This way of using it would help me improve my depth of thinking and understanding.*" Feedback on submitted assignments is essential for students' learning, but it also puts a lot of pressure on teachers, especially with a large number of students. In this case, GenAI may be a solution.

Moreover, AI can also provide learning resources tailored to students' specific needs. For example, a student majoring in English proposed an innovative learning methods, using ChatGPT to learn a second language, "*ChatGPT can generate short texts based on the words entered by the user to help students memorize the words.*" Moreover, some students from the Faculty of Education also assumed that AI can assist them in future teaching, e.g., "*I believe that the use of ChatGPT will reduce teachers' workload for answering questions. I may also use it to generate some lesson plans.*" Since GenAI was considered to "*improve students' motivation*" and "*help students learn better on their own*", in the future, it may potentially revolutionize traditional teaching and learning methods.

**(2) Writing and brainstorming support**

GenAI technologies, such as ChatGPT, can also be used as writing assistants. Sometimes, students find it difficult to generate ideas or find inspiration. In such cases, a participant suggested, "*It'll be convenient to ask ChatGPT some general questions and even get inspired by it.*" By inputting a question related to the writing topic, the AI output can serve as a starting point to develop and expand on their ideas. In addition, this virtual assistant is equipped to provide technical support, for example, "*it can help with formatting and information retrieval* "or "*help gather citations.*", which improves efficiency.

Furthermore, after writing, students can also use GenAI to enhance their writing skills. As one remarked, "*I would use it to help improve my writing (grammar, paraphrasing...), consult some questions or let it give some feedback on my writing.*" Especially for non-native speakers or students who are struggling with writing, it can be particularly useful if AI can "*help polish articles*" and provide personalized feedback for their written texts.

**(3) Research and analysis support**

The role of GenAI technologies in research has also caught the attention of students. In terms of its ability to acquire, compile, and consolidate information, some participants suggested it can "*facilitate literature searching,*" "*summarise readings,*" and "*even generate hypotheses based on data analysis.*" With a vast amount of data and knowledge, AI-powered technologies can help researchers always stay up-to-date with the latest research trends. Moreover, it also contributes to data collection and analysis. A student noted, "*It saves resources in data collection and initial analysis. We should ride on the initial insights to build our own insights.*" Since GenAI technologies are capable of rapidly and effectively processing large amounts of data, students can directly work on the basis of the preliminary analysis results.

**(4) Visual and audio multi-media support**

In addition to the above-mentioned uses, participants also used GenAI technologies for creating artworks and handling repetitive tasks. With advances in computer vision, AI-generated artworks have particularly gained attention from STEM students. A student from the faculty of science mentioned, "*I mainly played around with DALL-E, stable diffusion and other AI art technologies, which generate images based on a prompt.*" Similarly, an engineering student "*used text-to-image generation AI like stable diffusion at home to create artwork.*" Furthermore, AI technologies *can facilitate* "*the production of multi-media*", including slides, audios, and videos. As a content creator, "*when we have no clue how to visualize stuff, it can offer samples*

*and insights.*"

**(5) Administrative support**

Concerning "*repetitive or non-creative*" tasks, some participants believe that AI will perform well. As one commented, "*tedious administrative work will be handled by AI efficiently.*" By accelerating routine repetitive tasks, AI may leave more time for students to focus on their studies and research.

***What are the reasons behind students' concerns or lack of concerns regarding generative AI technologies?***

In alignment with the quantitative results, the qualitative data similarly revealed different concerns regarding challenges about GenAI. Some participants were optimistic about AI's integration in the future. The reasons for this optimism include the willingness mentioned earlier, as well as the belief that GenAI is part of the evolution and trends of technology. One student stated, "*It follows a general revolution of technology, similar to the public use of computers 40 years ago. I'm very much looking forward to the future of how such technology can reshape the world*" They suggested that as new technologies emerge, it is better to "*positively embrace it*" rather than avoid them.

Another reason behind the optimism was the assumption that humans would still maintain control and oversight over the GenAI. A participant remarked that "*I am not that concerned, as it would lead humans to smartly utilize such AI tools to complete their tasks in an efficient manner rather than simply being replaced by such tools.*" Another postgraduate student from the Faculty of Arts emphasized that AI is not a replacement for human skills and expertise: "*to my best of knowledge, I feel ChatGPT has not yet had the creativity and imagination as human beings, nor can it create a thesis for postgraduate students.*" At least for now, they believed humans would continue to be in the loop and have oversight over the GenAI technologies.

However, more than half of the participants still have concerns about the challenges of integrating GenAI technologies, mainly about the reliability of the technology itself and its impact:

**(1) Challenges concerning accuracy and transparency**

Currently, GenAI can promptly provide fluent and human-sounding responses, but their accuracy cannot always be guaranteed. As one student pointed out, "*We cannot predict or accurately verify the accuracy or validity of AI generated information. Some people may be misled by false information.*" Transparency is another significant concern. For a majority of users, the AI system is complex and opaque, which makes it difficult to understand how AI comes up with its decisions. "*It is always dangerous to use things you cannot understand*", a student noted. As AI-driven conversations become increasingly popular, remaining a "black box" may become an obstacle to public trust.

**(2) Challenges concerning privacy and ethical issues**

The use of GenAI also raised privacy and ethical concerns, which was mostly mentioned by students majored in arts and social science. They were worried that AI would collect personal information from our messages. As a social science student put forward, "*AI technologies are too strong so that they can obtain our private information easily.*" Since these messages will be used to further improve the system, if they are not properly protected, it "*can pose privacy and security risks.*"

Ethically, the plagiarism concern has been mentioned numerous times. Plagiarism has long been a critical issue in academics. But, with the rapid development of GenAI technologies, it has become increasingly difficult to identify plagiarized information. As an art student remarked, "*I want to know whether I am dealing with an AI bot or AI-generated content. Right now, it is somewhat easy to detect, but as the technology improves, it may not be so easy*"

**(3) Challenges concerning holistic competencies**

Regarding its impact on individuals and personal development, one of the main issues is over-reliance on AI, which may hinder people's growth, skills, and intellectual development over time. As one participant commented, "*this may lead to a decrease in critical thinking and make decisions only based on the information that AI provides to them.*" In addition to critical thinking, a student also noted its negative impact on creativity, "*some people may rely too much on AI technologies to generate ideas causing them to lose the capacity or willingness to think by themselves.*"

**(4) Challenges concerning career prospects**

Regarding its impact on society as a whole, GenAI also carries risks and drawbacks. The most frequently mentioned concern is job replacement. As GenAI is transforming the workplace, some jobs that students are preparing for may disappear. A computer science student expressed his concern "*I will probably lose my job in the future due to the advent of ChatGPT.*" Similarly, a student who majored in social science also mentioned, "*AI may replace the job that I'm interested in (e.g., GIS analyst)*". Consequently, employers may also raise their recruitment requirements. This development will pose a test for future graduates, since "*those who fall behind on this might have difficulty finding employment or catching up.*"

**(5) Challenges concerning human values**

Another mentioned societal risk relates to the value system. Some participants were worried that "AI could misalign with our human values and becomes a danger to us." For example, it may contribute to social injustice and inequality, as some participants noted, "*it may widen the gap between the rich and the poor*" and "*also be unfair to those students who don't use it.*" Furthermore, in academic institutions and education, some were concerned that the widespread use of AI might also might affect the student-teacher relationship, since students may be "*disappointed and lose respect for teachers.*"

**Discussion**

The study of student perceptions of GenAI, such as ChatGPT, in higher education reveals a complex and nuanced picture of both enthusiasm and concerns. The findings of this study provide an insightful understanding of university students' perception. It is evident that students are generally familiar with GenAI technologies, and their level of familiarity is influenced by factors such as knowledge about GenAI and frequency of use. The results also highlight the potential benefits and risks associated with using GenAI in teaching and learning, which are perceived differently among students based on their experiences with GenAI technologies. Overall, the participants showed a good understanding of the capabilities and limitations of GenAI technologies, as well as a positive attitude towards using these technologies in their learning, research, and future careers. However, there were also concerns about the reliability, privacy, and ethical issues associated with GenAI, as well as its potential impact on personal development, career prospects, and societal values. Table 5 shows the benefits and concerns of employing GenAI technologies.

The study revealed that students' knowledge of GenAI technologies and frequency of use are positively correlated. This suggests that exposure to these technologies and hands-on experience may help in enhancing students' understanding and acceptance of GenAI. Also, despite the relative novelty of GenAI for public use, students appear to have knowledge of the technologies and understand its benefits and risks quite well. Both quantitative and qualitative findings also show that students are generally willing to use GenAI for their studies and future work, but they have high expectations. For example, the study found that students perceive GenAI technologies as beneficial for providing personalized learning support as they expect learning resources tailored to their needs 24/7. In terms of writing and brainstorming support, students want feedback to improve writing skills, beyond just grammar checking and brainstorming, similar to the findings in Atalas' study (2023). For research and analysis support, students envision GenAI capabilities to not only facilitate literature searching and summarizing readings but also to generate hypotheses based on data analysis, enabling them to stay up-to-date with the latest research trends and build upon initial insights for their own work (Berg, 2023) which would not be expected from previous educational technologies. These findings indicate the potential of GenAI in revolutionizing traditional teaching and learning methods by offering tailored assistance, diverse learning needs, promoting efficiency and fostering self-directed learning.

**Student Perception of GenAI Technologies**

| Benefits related to | Challenges concerning |
|---|---|
| 1. Personalized and immediate learning support | 1. Accuracy and transparency |
| 2. Writing and brainstorming support | 2. Privacy and ethical issues |
| 3. Research and analysis support | 3. Holistic competencies |
| 4. Visual and audio multi-media support | 4. Career prospects |
| 5. Administrative support | 5. Human values |

Table 5 Benefits and Challenges on Generative AI Technologies from Student Perception

Despite the positive outlook, the study also reveals challenges concerning GenAI technologies, with students expressing reservations about over-reliance on the technology, its potential impact on the value of university education, and issues related to accuracy, transparency, privacy, and ethics. Students express concerns about the accuracy and ethical issues, particularly plagiarism, as they face difficulty in determining the originality of work generated by GenAI tools (Peres et al., 2023), which are unable to assess validity or identify falsehoods, thus necessitating human oversight (Lubowitz, 2023). Interestingly, there is no significant correlation between students' concerns and their knowledge about GenAI technologies, suggesting that even those with a good understanding of the technology may still have reservations, similar to Dahmash et al. (2020)'s findings. Additionally, students were apprehensive about GenAI, which may hinder critical thinking and creativity, and the impact of GenAI on job prospects (Gong et al., 2019; Ghotbi et al., 2022; Park et al., 2020) and human values (Gillissen et al., 2022; Jha et al., 2022).

User acceptance is key to the successful uptake of technological innovations, and students are the primary users of educational technologies. By understanding how students perceive generative AI technologies, educators and policymakers can better understand how best to integrate these technologies into higher education to enhance teaching and learning outcomes.

As mentioned, the reasons behind students' willingness and concerns about GenAI technologies are multifaceted. On one hand, students are optimistic about the future integration of these technologies into their academic and professional lives, considering GenAI as part of the ongoing technological evolution. On the other hand, students have reservations.

**Conclusion**

In this study, student perception of GenAI technologies were investigated. According to Biggs (1999; 2011), student perceptions of their learning environment, abilities, and teaching strategies significantly influence their learning approach and outcomes, with positive perceptions leading to a deep learning approach and negative perceptions resulting in a surface approach. Thus, it is vital to understand student perception in the context of GenAI technologies. By taking students' perceptions into account, educators and policymakers can better tailor GenAI technologies to address students' needs and concerns while promoting effective learning outcomes.

Understanding students on their willingness and concerns regarding the use of GenAI tools can help educators to better integrate these technologies into the learning process, ensuring they complement and enhance traditional teaching methods. This integration can lead to improved learning outcomes, as students will be more likely to adopt a deep approach to learning when they perceive GenAI as a valuable and supportive resource. Students' perceptions can provide insights into their level of AI literacy, which is essential for responsible use of GenAI technologies. By identifying gaps in students' understanding, educators can develop targeted interventions to improve AI literacy and prepare students for future employment in an increasingly AI-driven world. In the findings, students highlight the potential risks and concerns, educators can create guidelines and safeguards that ensure responsible and ethical use of GenAI technologies.

**Implications**

The diverse range of opinions among the participants highlights some implications that must be considered to ensure the successful integration of GenAI into higher education. Firstly, institutions should consider providing educational resources and workshops to familiarize students with GenAI technologies and their ethical and societal implications. This would enable students to make informed decisions when using these technologies in their academic endeavors.

Secondly, the development and implementation of GenAI technologies should prioritize transparency, accuracy, and privacy to foster trust and mitigate potential risks. For example, technical staff could work on explainable AI models that provide clear explanations of their decision-making processes. In addition, robust data protection policies and practices should be in place to safeguard users' privacy.

Lastly, higher education institutions should consider rethinking their policy, curricula and teaching approaches to better prepare students for a future where GenAI technologies are prevalent. This may involve fostering interdisciplinary learning, emphasizing critical thinking and creativity, and cultivating digital literacy and AI ethics education.

In conclusion, this study sheds light on the diverse perspectives of university students towards GenAI technologies and underscores the need for a balanced approach to integrating these technologies into higher education. By addressing students' concerns and maximizing the potential benefits, higher education institutions can harness the power of GenAI to enhance teaching and learning outcomes while also preparing students for the future workforce in the AI-era.

**Limitations and Future Research**

This study has several limitations that should be considered when interpreting the findings. First, the sample size was relatively small, which may limit the generalizability of the results to the broader population of students in Hong Kong. The study's reliance on self-reported data may also introduce potential biases, as participants could have been influenced by social desirability or inaccurate recall of their experiences with GenAI technologies. Furthermore, the cross-sectional design of the study does not allow for an examination of changes in students' perceptions over time as their exposure to and experiences with GenAI technologies evolve. Lastly, the study did not directly explore the actual impact of GenAI on students' learning outcomes, which would be necessary to provide a more comprehensive understanding of the role of these technologies in education.

Future research should address these limitations by employing larger, more diverse samples; using longitudinal designs to track changes in students' perceptions of generative AI over time and explore how these technologies are integrated into higher education; and examining the relationship between GenAI use and learning outcomes. Additionally, future research could explore on a specific group of students from different discipline, academic backgrounds, age groups, or cultural contexts on AI literacy.

Overall, there is a need for further research to better understand how best to integrate generative AI into higher education while minimizing potential risks related to privacy and security. By exploring these areas, we can ensure that these technologies are used responsibly and effectively in teaching and learning contexts.

**Declarations:**

Availability of data and material: The datasets used and/or analysed during the current study are available from the corresponding author on reasonable request

Acknowledgements: The author wishes to thank the students who participated the survey.